\documentclass[10pt, a4paper, twocolumn]{article}

\usepackage[dvips]{epsfig}
\usepackage{cite}
\usepackage{amssymb}
    \newcommand{\figwidth}{0.9\columnwidth}

\begin{document}
\twocolumn[
\title{Spin-resonance modes of the spin-gap magnet TlCuCl$_3$.}

    \author{V. N. Glazkov, A. I. Smirnov\\
        \textit{\small P. L. Kapitza Institute for Physical Problems RAS, 117334 Moscow,
        Russia}\\
    H.Tanaka, A.Oosawa\\
        \textit{\small Department of Physics, Tokyo
        Institute of Technology, Meguro-ku, Tokyo 152-8551, Japan }}

%\date{\today}

\maketitle

 \textbf{Abstract:} Three kinds of magnetic resonance
signals were detected in crystals of the spin-gap magnet
TlCuCl$_3$. First, we have observed the microwave absorption due
to the excitation of the transitions between the singlet ground
state and the excited triplet states. This mode has the linear
frequency-field dependence corresponding to the previously known
value of the zero-field spin-gap of 156 GHz and to the closing of
spin-gap at the magnetic field $H_c \approx$ 50 kOe. Second, the
thermally activated resonance absorption due to the transitions
between the spin sublevels of the triplet excitations was found.
These sublevels are split by the crystal field and external
magnetic field. Finally, we have observed antiferromagnetic
resonance absorption in the field-induced antiferromagnetic phase
above the critical field $H_c$. This resonance frequency is
strongly anisotropic with respect to the direction of the magnetic
field. \vspace*{5mm}]

\section{Introduction.}

The spin-gap magnets have been intensively studied during last
decades because of various quantum-disordered states found there.
The spin-gap structures were found in one-dimensional
antiferromagnets like dimerised spin $S=1/2$ chains, including
spin-Peierls magnets \cite{Hase}, spin $S=1$ chains
\cite{Regnault32}, spin ladders \cite{Dagotto} and dimer
structures \cite{Kageyama}. While the disordered ground states of
spin-gap systems are stable with respect to a weak interchain
exchange or anisotropy, they demonstrate an antiferromagnetic
ordering induced by  weak doping \cite{Renard} or by a strong
magnetic field \cite{oosawa-jpcm}. Impurities destroy locally the
spin-gap state and restore a local antiferromagnetic order around
impurity atoms, the overlapping of these area of local ordering
and a weak interchain exchange provide a long range magnetic
order. Magnetic field closes the spin gap, thus an ordered state
becomes possible. The spin gap magnet TlCuCl$_3$ is a unique
substance demonstrating both impurity-induced ordering in zero
field \cite{oosava-impurities} and field induced ordering in pure
crystals \cite{oosawa-jpcm, tanaka-order-neutrons}.

The crystals of  TlCuCl$_3$ have a dimer spin structure formed by
the $S=1/2$ spins of Cu$^{2+}$ ions. The dimers construct infinite
double spin chains coupled to each other, thus the system is
strongly coupled 3D dimer network, the structure of exchange
interactions is described in \cite{excitations}. The strongest
antiferromagnetic exchange is within the chemical dimer
Cu$_2$Cl$_6$ (5.68 meV), the spin-gap value is 0.65 meV.

 The field-induced antiferromagnetically ordered phase was found in
a magnetic field  $H>H_c\sim50$ kOe \cite{oosawa-jpcm} in
measurements of the magnetization. Formation of the field-induced
long-range magnetic order was confirmed by means of
neutron-scattering experiments \cite{tanaka-order-neutrons}. The
transition field has a strong temperature dependence
\cite{oosawa-jpcm,oosawa-specheat}. This dependence, unexpected in
the mean-field theory, was qualitatively described in the model of
Bose-Einstein condensation of magnons \cite{Nikuni}.

High-frequency magnetic resonance measurements
\cite{tanaka-esr,takatsu-esr} have demonstrated directly  the
field dependence of the energy gap, but were limited to the fields
$H<H_c$ and a few number of microwave frequencies.   Measurements
of zero momentum magnetic excitation energy in a wide range of the
magnetic fields were performed in the neutron scattering
investigation \cite{Ruegg}. The closing of the spin gap in the
magnetic field $H=H_c$ was confirmed, but the low-energy range of
antiferromagnetic resonance frequencies is beyond the resolution
of this experiment because of the strong field-induced magnetic
Bragg contamination below 0.75 meV (176 GHz). Thus, no
low-frequency response was found in the field range above $H_c$,
and  it was suggested that the energy of the lowest excitations is
zero at $H>H_c$. In the present paper we describe the detailed
magnetic resonance study of the crystal samples of TlCuCl$_3$ in
the range of microwave  frequencies 9 --- 80 GHz, in magnetic
fields up to 80 kOe. We detected at the first time the magnetic
resonance signals in the field-induced ordered phase and measured
their frequency-field dependences, which appeared to be nonlinear
and strongly anisotropic with respect to the direction of the
magnetic field. Besides, we observed the microwave absorption due
to the transitions from the ground state to the excited triplet
states. Finally, we found electron spin resonance (ESR) signals of
the thermally excited triplet excitations. This kind of resonance
is due to the transitions between the  spin sublevels of triplet
states. The evolution of triplet split spectrum to a single
exchange narrowed line was observed.

\section{Samples and experimental details.}

The crystals of TlCuCl$_3$ have monoclinic symmetry with the space
group $C_{2h}^5$ ($P2_1/c$). The two-fold axis is denoted as $b$.
The axes $a$ and $c$ form an angle of 96.32$^{\circ}$.

The sample growth is described in details in
Ref.\cite{oosawa-jpcm}. Crystals  have cleavage planes $(010)$ and
$(10\overline{2})$. During our experiments we have mounted
crystals in the following orientations with respect to the
magnetic field: $H||[010]$ ($b$-axis), $H\perp(10\overline{2})$
and $H||[201]$. The $[201]$ direction forms an angle of
51.4$^{\circ}$ with the $a$ axis. We used single crystals with the
volume of about $\sim$20\ldots50mm$^3$. The crystals are
hygroscopic, and a hydrated phase was present on the surface of
the samples, giving a parasitic paramagnetic resonance signal,
this signal grew when samples were exposed at open atmosphere.

ESR spectra were taken by use of a set of home made microwave
spectrometers with transmission type cavities and a
superconducting magnet.  The temperature of the sample was
regulated by pumping the vapour above the helium bath surrounding
the resonator or, in another cell, by heating the resonator within
the vacuum chamber filled with a small amount of heat exchanging
gas. ESR spectra were recorded as field dependences of the
microwave power transmitted through the cavity with the sample.

\section{Experimental results.}

\subsection{ESR at different temperatures.}
\begin{figure}
  \centering
  \epsfig{file=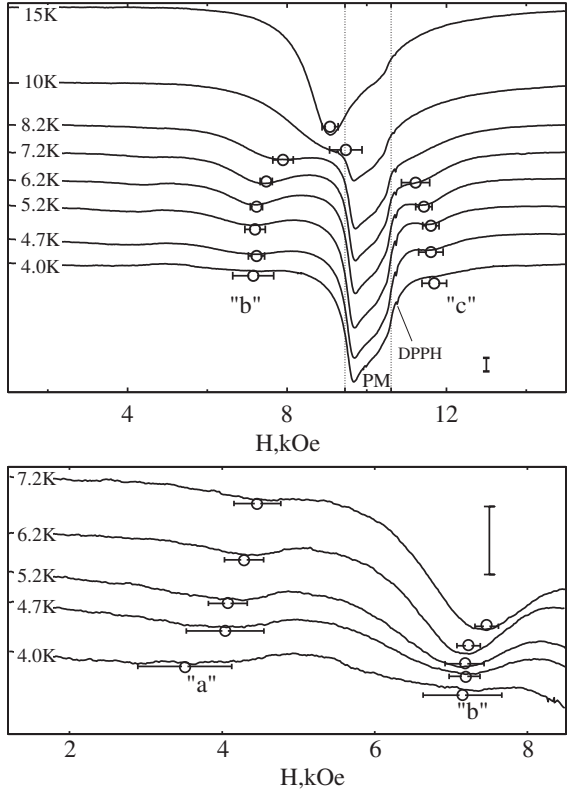, width=\figwidth, clip=}
  \caption{Temperature evolution of the low-field part of the ESR spectrum at
  $\mathbf{H}\perp(10\overline{2})$ and $f$=30.05 GHz. Lower panel represents
  a blowup
  of the data (vertical bar on both panels has the same
  absolute value). Letters ``a'', ``b'', ``c'' denote thermally activated components of the ESR
  spectrum. Circles  with error bars mark values of the resonance
  fields for  components ``b'' and ``c''.
  Absorption signal marked as PM is due to the parasitic hydratation of the sample surface.
  The narrow line on the right wing of the PM signal is the DPPH-mark ($g$=2.0).}
  \label{fig:scans(t)-tr}
\end{figure}

\begin{figure}
  \centering
  \epsfig{file=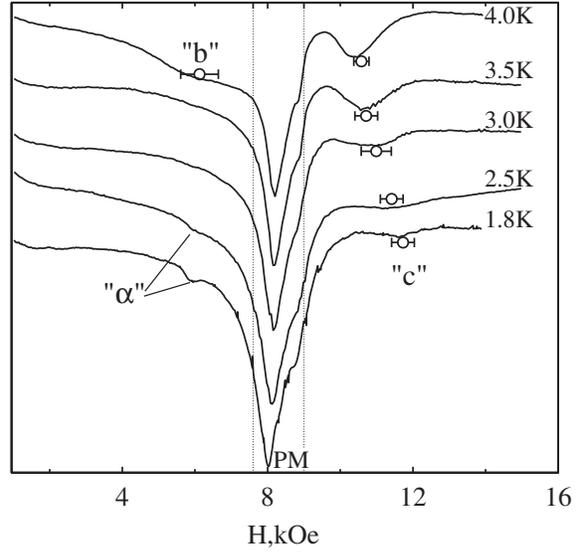, width=\figwidth, clip=}
  \caption{Evolution of the low-field ESR spectrum below  $T$=4.2 K at
    $\mathbf{H}||b$, f=25.94GHz. ``b'' and ``c'' mark thermally activated ESR components,
   ``$\alpha$'' is a weak paramagnetic signal with $g\approx3$. Circles with error bars mark
 resonance fields for  components ``b'' and ``c''. }
  \label{fig:scans(t)-lt}
\end{figure}

\begin{figure}
  \centering
  \epsfig{file=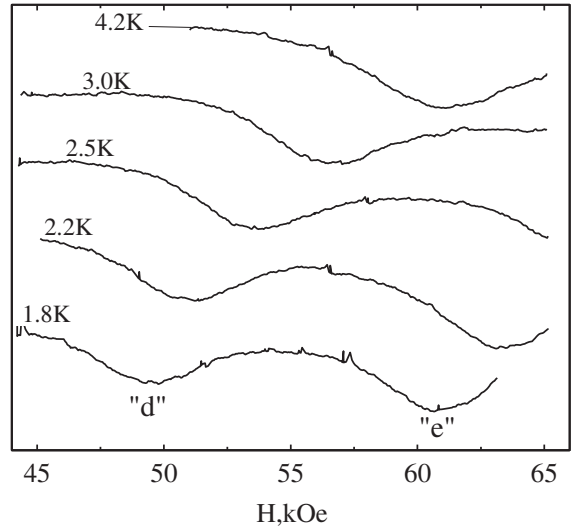, width=\figwidth, clip=}
  \caption{Temperature evolution of the high field ESR spectrum at $\mathbf{H}||b$, $f$=25.94GHz.
  Letters ``d'', ``e'' mark spectral components.}
  \label{fig:scans(t)-hf}
\end{figure}

\begin{figure}
  \centering
  \epsfig{file=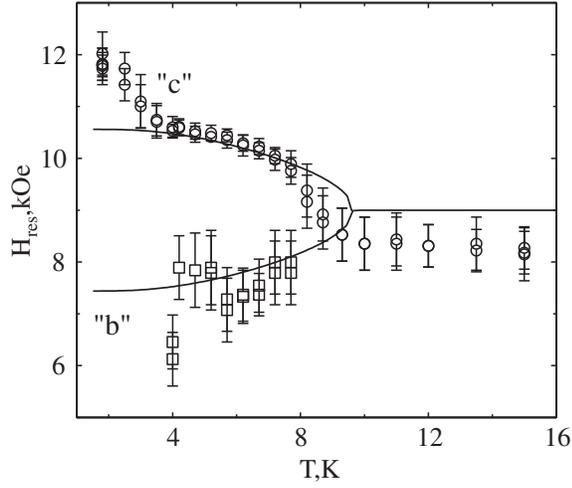,  width=\figwidth, clip=}
  \caption{Temperature dependences of the ESR fields of components ``b'' and ``c''
   at $\mathbf{H}||b$, f=25.94GHz.
   Solid curves represent exchange narrowing theory (see text).}
  \label{fig:fields(t)}
\end{figure}

\begin{figure}
  \centering
  \epsfig{file=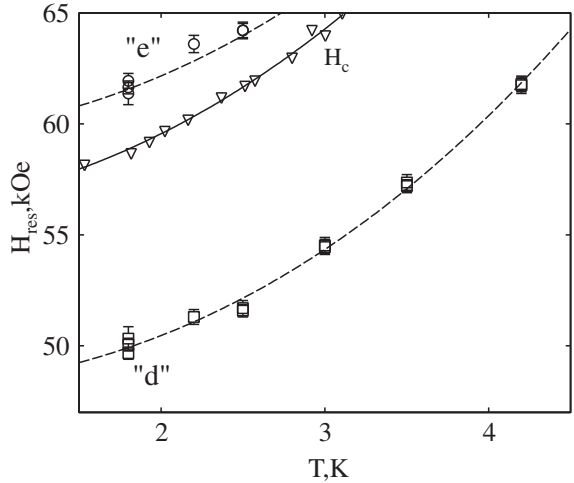,  width=\figwidth, clip=}
  \caption{Temperature dependences of the ESR fields for the high-field components ``d'' and ``e''
  at $\mathbf{H}||b$, f=25.94GHz.
   Circles --- component ``e'', squares --- component ``d'', triangles ---
   H$_c$(T) data from Ref. \cite{oosawa-specheat}.
   Curves are guide to the eye}
  \label{fig:fields(t)hf}
\end{figure}

\begin{figure}
  \centering
  \epsfig{file=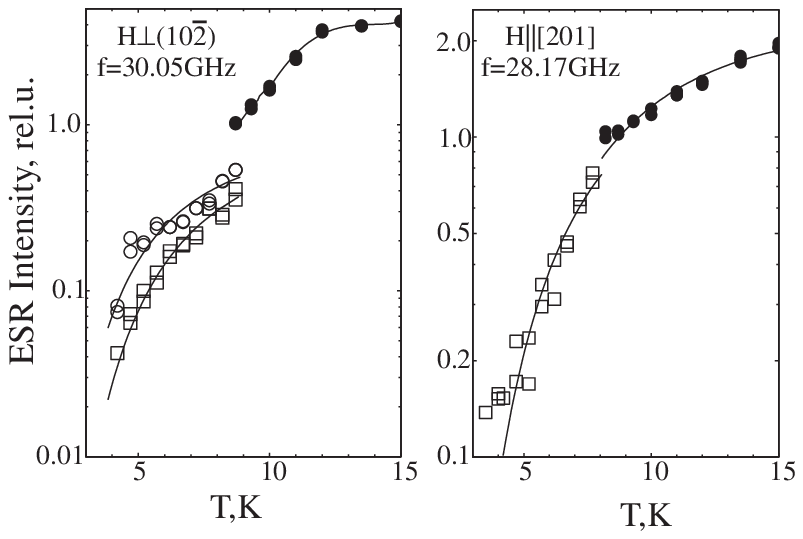, width=\figwidth,  clip=}
  \caption{Temperature dependence of the ESR integrated intensity.
  Symbols: filled circles --- exchange narrowed ESR line above the splitting temperature,
  open circles --- component ``b'', squares ---
  component ``c''.  Curves are guide to the eye.}
  \label{fig:int(t)}
\end{figure}

The temperature evolution of an ESR absorption curve in the
low-field range is presented in Figs. \ref{fig:scans(t)-tr} and
\ref{fig:scans(t)-lt}, the high-field data are illustrated in Fig.
\ref{fig:scans(t)-hf}. The absorption spectrum consists of several
components, which demonstrate different kinds of temperature
dependences of the intensity and the resonance field.

At first, there are several components located near the
paramagnetic resonance field (e.g., near 10 kOe  for the microwave
frequency $f$=30.05 GHz, as shown on Fig.\ref{fig:scans(t)-tr}).
The parasitic absorption takes place in the field range marked as
PM, and has the Curie-like behavior of the integrated intensity.
The shape of this absorption curve is typical for powder samples,
with two sharp boundaries originating from the maximum and minimum
values of an anisotropic $g$-factor. The lineshape and position of
this absorption do not depend on the orientation of the magnetic
field with respect to crystal axes. From the intensity of the
paramagnetic resonance signal of the hydrated surface we estimate
the number of the paramagnetic ions in this spoiled area of the
crystal as 0.02 from the total number of magnetic ions.

Apart from the parasitic signal, we observe thermally activated
resonance absorption (lines marked in order of their resonance
field increase as ``a'', ``b'' and ``c'', see Fig.
\ref{fig:scans(t)-tr}). The resonance fields of these components
are temperature dependent, they become closer to each other as
temperature rises and finally coalescence to a single line at
$T\approx 8$K. As expected for Cu$^{2+}$ ions, the $g$-factor of
this single line is close to 2.0 and its anisotropy does not
exceed 15\%. At low temperature the resonance field of the
component ``a'' is close to a half of the resonance field of free
spins $H_f=hf/g\mu_B H$ ($f$ is the microwave frequency), and the
resonance fields of components ``b'' and ``c'' are located on both
sides from $H_f$. The splitting of the single resonance line into
several components with decreasing temperature was observed for
all three orientations of the applied magnetic field. However, the
half-field absorption component ``a'' was not observed for
$\mathbf{H}||b$, and the component ``b'' can not be resolved on
the background of the strong parasitic paramagnetic absorption for
$\mathbf{H}||[201]$. The temperature dependences of resonance
fields and intensities for thermally activated components are
given in Figs. \ref{fig:fields(t)}, \ref{fig:int(t)}. The ``a''
component intensity is small, the most part of the ESR intensity
is concentrated in ``b'' and ``c'' components. The intensities of
these components are slightly different (this difference is of
about of 10\% of the total intensity). The position of the more
intensive component is orientation dependent: for
$\mathbf{H}\perp(10\overline{2})$ the ``b'' component is more
intensive, while for $\mathbf{H}||[201]$ the ``c'' component
conserves more then a half of the total intensity when measured at
the temperature of splitting. The intensities of components ``a'',
``b'', ``c'' decrease quickly with decreasing temperature,
practically disappearing near 1.5 K.

The  ESR absorption in the high-field range (Fig.
\ref{fig:scans(t)-hf}) was observed only at temperatures below 4.2
K. Here the absorption line consists of the two components marked
by the letters ``d'' and ``e'', located on both sides of the
critical field $H_c$.  Both components shift to the higher fields
with temperature increase (see Fig. \ref{fig:fields(t)hf}), in
accordance with the increase of the critical field $H_c$
\cite{oosawa-specheat}.

\subsection{Magnetic resonance spectra.}

\begin{figure}
  \centering
  \epsfig{file=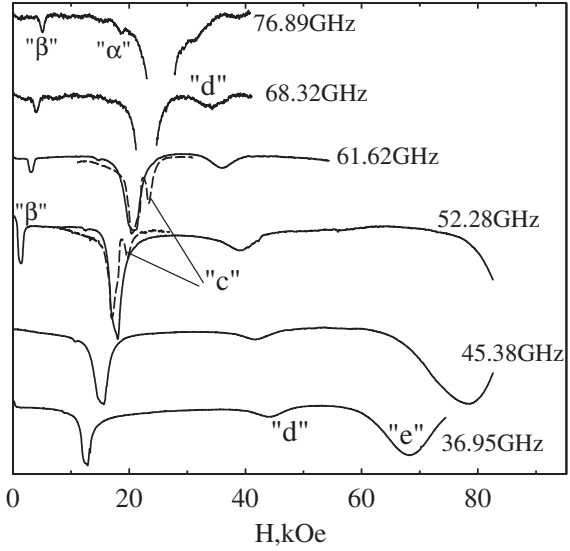,  width=\figwidth,  clip=}
  \caption{ESR spectra at different frequencies for $\mathbf{H}||b$.
  Solid lines: T=1.5K, dashed lines: T=4.2K. Intensive signals of the paramagnetic absorption
  at f=76.89~GHz and f=68.32~GHz are partially removed. Letters mark different spectral components
  as described in the text.}

  \label{fig:scans(f)}

\end{figure}

\begin{figure}
  \centering
  \epsfig{file=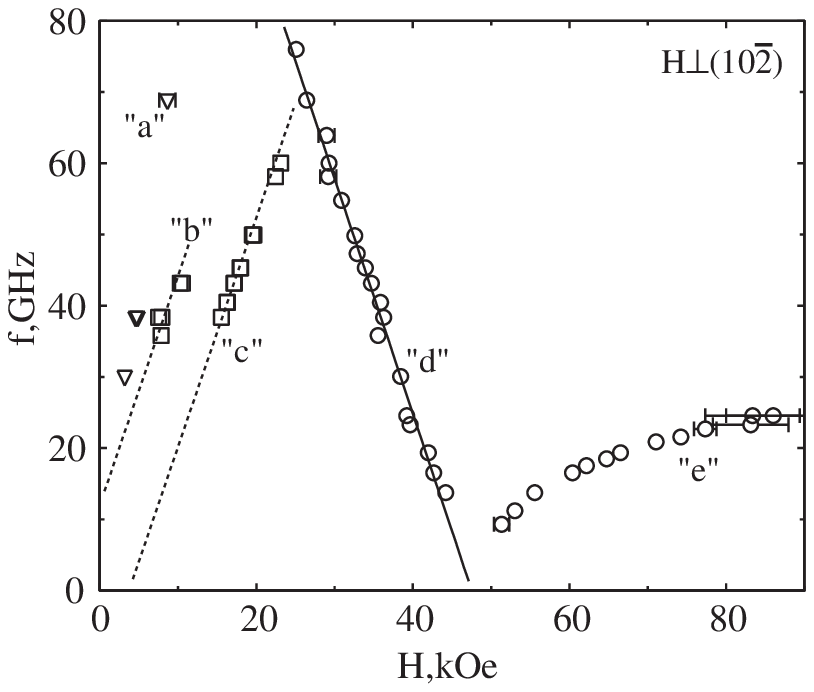,  width=\figwidth, clip=}
  \caption{Frequency-field dependences for spectral components
``a''---``e'' taken at $\mathbf{H}\perp(10\overline{2})$ and
$T$=1.5K.
  Lines are linear fits with parameters described in the text.}
  \label{fig:spectr1}
\end{figure}

\begin{figure}
  \centering
  \epsfig{file=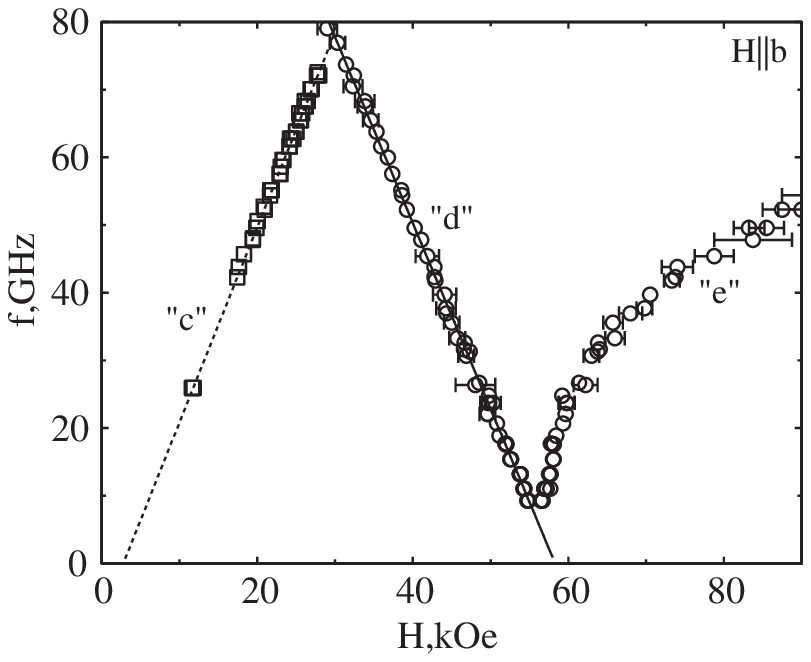,  width=\figwidth, clip=}
  \caption{
Frequency-field dependences for spectral components ``c''---``e''
taken at $\mathbf{H}||b$ and $T$=1.5K.
 Lines are linear fits with parameters described in the text.}
\label{fig:spectr2}
\end{figure}

\begin{figure}
  \centering
  \epsfig{file=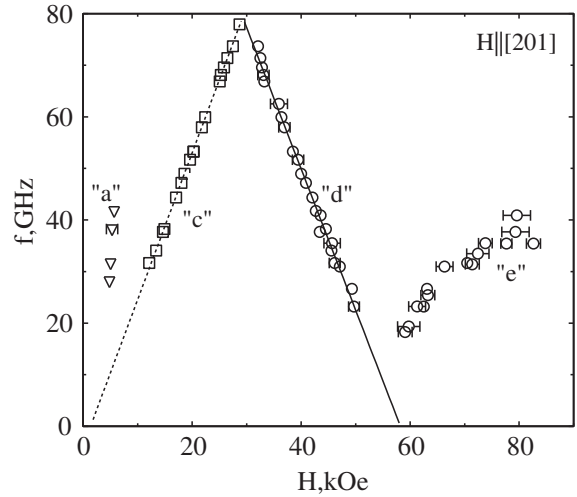,  width=\figwidth, clip=}
  \caption{
Frequency-field dependences for spectral components ``a'',
``c''---``e'' taken at $\mathbf{H}||[201]$ and $T$=1.5K.
 Lines are linear fits with parameters described in the text.
    }
  \label{fig:spectr3}
\end{figure}

ESR absorption lines taken at different frequencies are presented
at the Fig. \ref{fig:scans(f)}. Besides of the above-described
components ``a''\ldots``e'', two additional absorption components
(``$\alpha$'' and ``$\beta$'' in Fig. \ref{fig:scans(f)}) are
observed to the left of the paramagnetic resonance field. A weak
line ``$\alpha$'' demonstrates a linear spectrum with $g$-factor
value close to 3.0. It is probably due to impurities, their
concentration is of about $\sim10^{-4}$. The component ``$\beta$''
demonstrates nonlinear and anisotropic frequency-field dependence
with the gap of 50~GHz. Its shape is irregular and intensity
increases with decreasing temperature. The origin of the component
``$\beta$'' is unclear. Probably, it is also due to impurities or
additional parasitic phase.

The measured field-dependences of ESR frequencies for three
perpendicular orientations of the magnetic field ($\mathbf{H}||b$,
$\mathbf{H}\perp(10\overline{2})$ and $\mathbf{H}||[201]$) are
given in Figs. \ref{fig:spectr1}, \ref{fig:spectr2} and
\ref{fig:spectr3}.

\begin{figure}
  \centering
  \epsfig{file=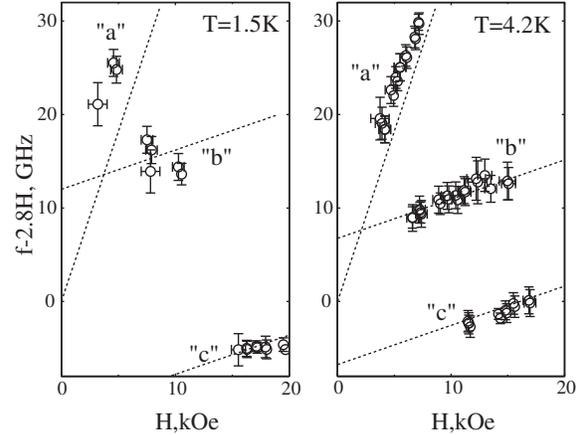, width=\figwidth,   clip=}
  \caption{ESR spectra of the triplet excitations at
  $\mathbf{H}\perp(10\overline{2})$,
  T=1.5K (left) and T=4.2K (right).
  The linear contribution, corresponding to the $g$=2.0 ESR is subtracted. Lines
  present the linear approximation (see text).
  }\label{fig:spectra-tr102}
\end{figure}

The resonance fields of  thermally activated components ``a'',
``b'' and ``c'' increase with the increase of the microwave
frequency. ESR lines ``b'' and ``c'' show linear $f(H)$
dependences. However, the linear extrapolation of their resonance
frequencies to zero field results in nonzero values. The magnitude
of the splitting between ``b'' and ``c'' components  is of about
10~GHz. The fine details of the frequency-field dependences of
thermally activated components measured at two temperatures 4.2 K
and 1.5 K are shown in the Fig. \ref{fig:spectra-tr102}. Note,
that for the observation of the ESR signals from noninteracting
triplet excitations (at small population numbers) we have to deal
at the lowest temperature. From the other side the freezing out of
the intensity of these components complicates the detection of
signals and the measurement of resonance fields. At both
temperatures the observed ESR frequencies of the ``b'' and ``c''
spectral components may be described well by the linear equation

\begin{equation}\label{eqn:split-fit-eqn}
    f=\frac{g\mu_BH}{h}\pm A
\end{equation}

The fitting procedure according to equation
(\ref{eqn:split-fit-eqn}) for temperature independent but
anisotropic  $g$-factor, and temperature-dependent constants $A$
results in the following values:

\begin{itemize}
    \item[]$\mathbf{H}\perp(10\overline{2})$: $g=2.30\pm0.05$, $A(4.2K)=6.8\pm1.0$GHz, $A(1.5K)=12\pm2$GHz
    \item[]$\mathbf{H}||b$: $g=2.06\pm0.02$, $A(4.2K)=4.5\pm0.5$GHz, $A(1.5K)=8.0\pm0.7$GHz
    \item[]$\mathbf{H}||[201]$: $g=2.03\pm0.015$, $A(4.2K)=2.6\pm0.4$GHz, $A(1.5K)=3.9\pm0.5$GHz
\end{itemize}

The correspondence of the fitting dependences and experimentally
observed frequencies is illustrated on the
Fig.\ref{fig:spectra-tr102} and is quite satisfactory. The
$g$-factor values determined from the fitting of the split
low-temperature spectra coincides within the experimental error
with the $g$ values measured above 10K, as well as with the
$g$-factor values measured in the Ref.\cite{oosawa-jpcm}
($g_b$=2.06, $g_{\perp(10\overline{2})}$=2.23).

High-field absorption components ``d''  and ``e''  differ strongly
in their frequency-field dependences. The resonance ``d'' (to the
left of the critical field $H_c$) linearly shifts to lower fields
at the frequency increase. This ESR mode can be identified as
absorption due to the transitions between singlet ground state and
gapped triplet states, analogous to the absorption observed in
\cite{tanaka-esr,takatsu-esr}. The component ``e'' shifts to the
higher fields with increasing frequency, its frequency-field
dependence is nonlinear and anisotropic with respect to the
orientation of the external field. This resonance mode was not
reported in previous investigations.

\section{Discussion.}
\subsection{ESR of the triplet excitations.}

The thermal activation of the absorption lines ``a'', ``b'', ``c''
indicates that they are due to the gapped triplet excitations. The
growth of intensity  of these components beginning at $T\simeq4$~K
correlates well with the growth of the susceptibility measured in
Ref.\cite{oosawa-jpcm} and with the gap value 7.7 K. A remarkable
feature of the observed triplet ESR is the splitting of this
resonance into two components, and arising of a weak component
``a''.

While the magnetic resonance frequency of isolated $S=1/2$
Cu$^{2+}$ ions is not affected by the crystal field, the
excitations of a system of exchange-coupled Cu$^{2+}$ ions are
triplets carrying spins $S=1$. Therefore the magnetic resonance
frequency of triplets should be influenced by a crystal field. The
crystal field splits magnetic sublevels of a spin $S$=1 state even
at zero magnetic field \cite{AbragamBleaney}. This splitting
results in a three-component spectrum, corresponding to the
transitions between different pairs of the spin sublevels. Thus,
we interpret the observed lines ``a'', ``b'', ``c'' as  absorption
due to transitions between the sublevels of triplet excitations,
inegrated over the thermally excited ensemble of triplet
excitations.

The coalescence of the ESR line components at the temperature
increase corresponds to the scenario of the formation of the
exchange narrowed ESR line \cite{anderson-stat}: when the exchange
interaction between the spins precessing with different
frequencies is weak (slow exchange limit), there are separate
modes of magnetic resonance. For the opposite limit, when the
exchange frequency is greater than the difference in
eigenfrequencies,  a single collective mode is formed, known as
exchange-narrowed mode. For thermally activated spins the exchange
frequency is temperature dependent. The scenario with the
transition from the slow to the fast limit of the exchange
narrowing of ESR lines was observed earlier, e.g., in magnetic
resonance study of molecules with thermally activated spin states
\cite{chestnut} and at the temperature evolution of the ESR
spectra of the spin-Peierls magnet CuGeO$_3$ doped with impurities
\cite{GlazkovJETP,JETPL}. The observed evolution of the triplet
ESR lines in TlCuCl$_3$ follows the same scenario. At low
temperatures the concentration of triplets is small and
three-component ESR of noninteracting spins $S=1$ in a crystal
field is observed. At the larger concentration of activated spins
(at higher temperatures) the components merge in a single ESR
line.

Starting from this qualitative consideration of thermally
activated resonances we can further describe on a quantitative
level the ESR frequencies  at low temperature and the temperature
evolution.

The ESR frequencies are explained (see e.g. Ref.\cite{DateKindo})
by assuming an effective Hamiltonian for the triplet with the
wavevector $\mathbf{k}$ given by

\begin{equation}\label{eqn:aniz-ham}
  {\cal{H}}_{\mathbf{k}}=\mu_B\mathbf{S}\widehat{g}\mathbf{H}+D_{\mathbf k}S_z^2+E_{\mathbf k}(S_x^2-S_y^2)
\end{equation}

here $\widehat{g}$ is a $g$-tensor, $D_{\mathbf{k}}$,
$E_{\mathbf{k}}$ are the anisotropy constants. The anisotropy
constants values may be ${\mathbf k}$-dependent, as, e.g.  in a
Haldane magnet Ref.\cite{Meshkov}.

 Considering the anisotropy terms as a
perturbation to the Zeeman term, in the first order of the
perturbations theory one  obtains for the energy levels

\begin{eqnarray}
    E_{\pm}=\pm g\mu_BH+\frac{D_{\bf
    k}}{2}\left((\mathbf{z}\cdot\mathbf{n})^2+1\right)+~~~~~~~~~~~~~~\nonumber\\
    +\frac{E_{\bf k}}{2}\left((\mathbf{x}\cdot\mathbf{n})^2-(\mathbf{y}\cdot\mathbf{n})^2\right)
    \label{eqn:correct-pm}\\
    E_0=    D_{\bf k}\left(1-(\mathbf{z}\cdot\mathbf{n})^2\right)-
    E_{\bf k}\left((\mathbf{x}\cdot\mathbf{n})^2-(\mathbf{y}\cdot\mathbf{n})^2\right)
    \label{eqn:correct-0}
\end{eqnarray}

here $\mathbf{n}$ is the unit vector parallel to the magnetic field.

The observable spectrum should  be determined by the anisotropy
constants $D_{\mathbf k}$, $E_{\mathbf k}$ averaged over the
ensemble of triplet excitations at given temperature .  Thus, the
values $D$ and $E$, resulting from the averaging of $D_{\bf k}$
and $E_{\bf k}$, turn out to be temperature dependent. The
 averaged transition frequencies are

\begin{eqnarray}
    \hbar\omega_{1,2}=g\mu_BH\pm~~~~~~~~~~~~~~~~~~~~~~~~~~~~~~~~~~~~~~~~~~\nonumber\\
    \pm
    \left[
    \frac{D}{2}\left(3(\mathbf{z}\cdot\mathbf{n})^2-1\right)+
    \frac{3E}{2}\left((\mathbf{x}\cdot\mathbf{n})^2-(\mathbf{y}\cdot\mathbf{n})^2\right)
    \right]
    \label{eqn:freq1}\\
\hbar\omega_{3}=2g\mu_BH~~~~~~~~~~~~~~~~~~~~~~~~~~~~~~~~~~~~~~~~~~~~~~\label{eqn:freq2}
\end{eqnarray}

Two of three resonance frequencies correspond to the observed
frequency-field  dependences presented by
Eqn.(\ref{eqn:split-fit-eqn}) with the splitting constant $A$
dependent on the orientation only. The  resonance field of the
third mode  should be equal to the one-half of the field
$H_f=hf/g\mu_B H$.  The observed ESR  lines ``a'', ``b'', ``c''
correspond well to spin resonance modes $\omega_{3}$,
$\omega_{1,2}$ . The deviation of the resonance frequencies $2\pi
\omega_{1,2}$ from the free spin resonance frequency $g\mu_BH/h$
for one of the three arbitrary mutually orthogonal directions of
the external field should be equal to the sum of the deviations
for two other directions. This relation is valid for the measured
deviations $A$, confirming that the observed splitting is due to
the crystal field.

The triplet excitations are  multi-spin objects, hence, the
symmetry of the effective crystal field  should be determined by
the crystal symmetry and not by the symmetry of the local
surrounding of the magnetic ions. Thus, one axis of the symmetry
of the effective field should be aligned along the  two-fold axis
$b$, we note it as $\mathbf{z}||b$. The other two axes lie within
the $a-c$ plane of the crystal. To determine the values of the
anisotropy constants $D$ and $E$ we suppose that the $\mathbf{x}$
axis is directed along the direction of spin-ordering at ${\bf
H}\parallel b$ (i.e. at an angle of 39$^\circ$ from the $a$-axis
in the $a-c$ plane)
 \cite{tanaka-order-neutrons}.

Then, using Eqn.(\ref{eqn:freq1}), and observed values of $A$, we
obtain that $D$ and $E$ are of the same sign, and their magnitudes
are

\vspace{3mm}
\begin{tabular}{cl}
    at T=1.5K:& $|D/h|=8.0\pm0.7$GHz,\\
    &$|E/h|=5.8\pm0.6$GHz\\
    at T=4.2K:& $|D/h|=4.5\pm0.5$GHz,\\
    &$|E/h|=3.6\pm0.5$GHz
\end{tabular}
\vspace{3mm}

The signs of anisotropy constants can be determined from the
relation between the intensities of the spectral components shown
in Fig. \ref{fig:int(t)}. Due to the difference in population
numbers of different energy levels at finite temperature, at
$g\mu_BH \gg D,E$ the absorption related to the transitions
$|-\rangle\leftrightarrow|0\rangle$ is larger then that related to
the transitions $|+\rangle\leftrightarrow|0\rangle$. Position of
the resonance field $H_-$, corresponding to the transitions
$|-\rangle\leftrightarrow|0\rangle$, with respect to the field
$H_f$ depends on the orientation and on the sign of the anisotropy
constants (see Eqns.
(\ref{eqn:correct-pm}),(\ref{eqn:correct-0})):

\begin{eqnarray}
    H_-=H_f +\frac{1}{g\mu_B}\times~~~~~~~~~~~~~~~~~~~~~~~~~~~~~~~~~~~~~\nonumber\\
    \times \left[
    \frac{D}{2}\left(3(\mathbf{z}\cdot\mathbf{n})^2-1\right)+
    \frac{3E}{2}\left((\mathbf{x}\cdot\mathbf{n})^2-(\mathbf{y}\cdot\mathbf{n})^2\right)
    \right]
\end{eqnarray}

The more intensive component is observed to the left from $H_f$ at
$\mathbf{H}\perp(10\overline{2})$ and to the right from $H_f$ at
$\mathbf{H}||[201]$. Then we conclude that both $D$ and $E$ are
positive.

The observed change of the anisotropy constants $D$ and $E$ with
temperature  probably indicates the change of the distribution of
the thermally excited triplet excitations in the phase space. At
low temperature T=1.5 K  the triplets are excited mainly near the
bottom of the triplet band, while at T=4.2 K (this temperature
corresponds to one half of the spin-gap energy) the triplets are
excited over the whole band.

 The temperature evolution of the thermally activated
lines, shown in Fig. \ref{fig:fields(t)}, may be analyzed using
the approach of the exchange narrowed spin resonance
 \cite{anderson-stat}. We neglect here the intensity of the
almost prohibited two-quantum component ``a'', and the difference
between the intensities of components ``b'' and ``c''. The
frequency shift $\delta f$ from the center of gravity of the ESR
spectrum is given by the following equations of Ref.
\cite{anderson-stat}:

\begin{eqnarray}
    f_e>f_0&:&\delta f=0 \label{eqn:narrowing-start}\\
    f_e<f_0&:&\delta f=\pm\sqrt{f_0^2-f_e^2}
\end{eqnarray}

Here $\pm f_0$ are the deviations of the resonance frequencies
from the center of gravity of the ESR line in the absence of the
exchange narrowing, and $f_e$ is the average exchange frequency.

 For thermally excited magnetic states, the exchange
frequency is temperature dependent \cite{chestnut}

\begin{equation}\label{eqn:omegaex}
    f_e=F\exp\left(-\frac{\Delta}{k_BT}\right)
\end{equation}

here $\Delta=7.7$K is a zero-field energy gap.

The theoretical curves, calculated using
Eqns.(\ref{eqn:narrowing-start})-(\ref{eqn:omegaex}), are shown at
the Figure \ref{fig:fields(t)}. We have taken the splitting value
$f_0$ to be equal to splitting  at 4.2 K and used preexponential
factor $F$ as a fitting parameter. The fitting curves presented by
solid lines in Fig. \ref{fig:fields(t)} correspond  to the value
$F=10$ GHz.

The model of exchange narrowing used here is simplified. First, it
does not include the temperature dependence of the effective
anisotropy constants $D$ and $E$. Another simplification is
ignorance of the field dependence of the energy gap  and of the
different population numbers of energy levels. Particularly, this
simplified model can't explain the abrupt change of the resonance
frequency of component ``b'' between the temperatures 1.5 and 4 K.
Probably, this change is due to the expanding of the phase space
occupied by the excited triplets at $T>\Delta/2=4$ K, resulting in
change of the effective constants $D$ and  $E$.

\subsection{Singlet-triplet transitions.}

\begin{figure}
  \centering
  \epsfig{file=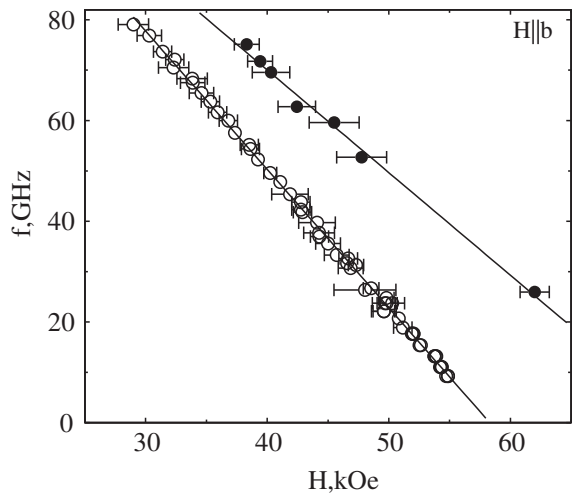, width=\figwidth,  clip=}
  \caption{ESR spectra of the component``d'' of the ESR spectrum. Open symbols ---
  T=1.5K, closed --- T=4.2K. $\mathbf{H}||b$.}\label{fig:spectra-str}
\end{figure}

 The ESR component ``d''  corresponds to direct over-gap
transitions between singlet ground state and gapped triplet
states. It is the same type of ESR absorption as reported earlier
for higher frequencies \cite{tanaka-esr,takatsu-esr}. The ESR
spectrum of this component may be described by the linear equation

\begin{equation}\label{eqn:s-tr-fit}
  f=\frac{1}{h}\left(\tilde{\Delta}-\tilde{g}\mu_BH\right)
\end{equation}

here $\tilde{\Delta}$ is expected to be close to the spin-gap
$\Delta$  and $\tilde{g}$ is the effective $g$-factor.

The  straight lines on Figs.\ref{fig:spectr1}, \ref{fig:spectr2}
and \ref{fig:spectra-str} present results of the linear fitting
according to  the relation (\ref{eqn:s-tr-fit}) with the
parameters given in Table \ref{tab:singlet-triplet}.

\begin{table}[h]
    \caption{\label{tab:singlet-triplet}}
    \centering
\begin{tabular}{|c|c|c|c|}
    \hline
     \multicolumn{2}{|c|}{}
     &$\tilde{g}$&$\tilde{\Delta}/h$, GHz\\
     \hline
     \hline
    T=1.5K&$\mathbf{H}||b$ & $1.96\pm0.15$&$160\pm10$ \\
    &$\mathbf{H}\perp(10\overline{2})$ & $2.36\pm0.15$&$157\pm10$\\
    &$\mathbf{H}||[201]$&  $1.96\pm0.15$&$159\pm10$\\
    \hline
    \hline
    T=4.2K&$\mathbf{H}||b$ & $1.5\pm0.1$&$154\pm6$\\
    &$\mathbf{H}\perp(10\overline{2})$ & $1.6\pm0.3$&$140\pm20$ \\
    &$\mathbf{H}||[201]$&$1.4\pm0.12$&$148\pm8$ \\
    \hline
\end{tabular}
\end{table}

The  value $\tilde{\Delta}$  obtained from linear extrapolation
corresponds well to the value of 156 GHz observed in  other ESR
\cite{tanaka-esr,takatsu-esr} and neutron studies\cite{Ruegg}. The
effect of crystal field discussed in the previous Section affects
the triplet levels and should distort the linear dependence
(\ref{eqn:s-tr-fit}): at zero field the  $f(H)$ curves taken at
different orientations of the magnetic field should start from the
same point $f(0)=\Delta$, then  at $g\mu_BH\gg D,E$ they should
have the linear asymptotic dependence, which may be easily
obtained from Eqn.(\ref{eqn:correct-pm}):

\begin{eqnarray}
\hbar\omega=\Delta-
g\mu_BH+~~~~~~~~~~~~~~~~~~~~~~~~~~~~~~~~~~~~~~~\nonumber\\
+\frac{D}{2}\left((\mathbf{z}\cdot\mathbf{n})^2+1\right)+
    \frac{E}{2}\left((\mathbf{x}\cdot\mathbf{n})^2-(\mathbf{y}\cdot\mathbf{n})^2\right)\label{eqn:singlet-triplet-anisotropy}
\end{eqnarray}

Thus, the measured values of $\tilde{\Delta}$ for different
orientations should differ from the spin-gap value $\Delta$ for
$\pm$5 GHz. This shift is within the experimental resolution for
the value of $\tilde{\Delta}$ obtained from the frequency of line
``d''.

The value of $\tilde{g}$ measured at $T=$ 1.5 K coincides well
with the $g$-factor values determined from the ESR of the
thermally activated triplets. However, at $T=$4.2 K the slope of
this linear dependence becomes smaller for the factor of 0.75,
while the value of $\tilde{\Delta}$ obtained from the
extrapolation to zero field remains constant within the
experimental error of 10\%. Thus the increase of the critical
field value at the increase of temperature should be ascribed to a
change of the field influence on the singlet-triplet gap with
temperature and not to the change of the zero-field gap.  It is
important to note here, that the value of $\tilde{g}$ is
temperature dependent, but the $g$-factor describing the
field-dependence of the sublevels of the excited triplets does not
depend on the temperature. The change of the effective $g$-factor
$\tilde{g}$  may be presumably attributed to the effect of the
thermal renormalization of magnon frequency observed for the
related compound KCuCl$_3$ \cite{Cavadini}. The temperature when
the value of $\tilde{g}$ is changing is about the spin-gap value
in the magnetic field, thus the renormalization is probable.

From the frequency-field dependences for singlet-triplet
absorption we  determine the critical field $H_c$  as a field, at
which the singlet-triplet transition frequency turns to zero.
These values measured at $T=1.5$ K are:
$H_c(\mathbf{H}||b)=58.3\pm1.0$kOe,
$H_c(\mathbf{H}\perp(10\overline{2}))=47.5\pm1.5$kOe and
$H_c(\mathbf{H}||[201])=58.1\pm2.0$kOe. These data are in
agreement with magnetization and specific heat data
\cite{oosawa-jpcm, oosawa-specheat}.

\subsection{Magnetic resonance in the field-induced magnetically ordered phase.}

\begin{figure}
  \centering
  \epsfig{file=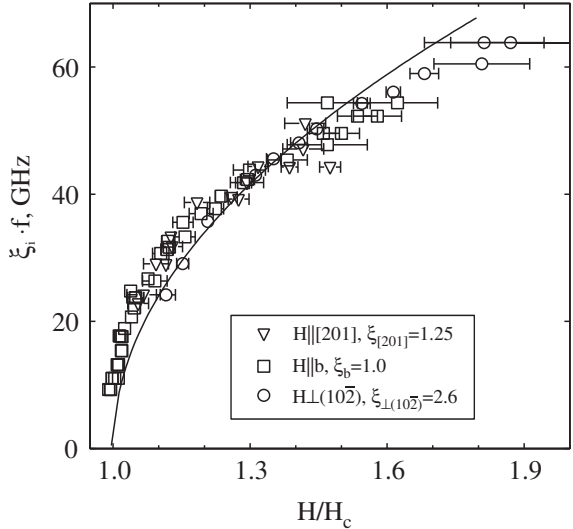, width=\figwidth,  clip=}
  \caption{Normalized frequency-field dependences of the antiferromagnetic
  resonance for three directions of the magnetic field.
    Solid curve is the fitting result (see text).}
  \label{fig:afmr-scale}
\end{figure}

The  component ``e'' of the ESR line observed at $H>H_c$, shown in
Fig. \ref{fig:scans(t)-hf}, demonstrates nonlinear and anisotropic
spectrum presented on Figs.
\ref{fig:spectr1},\ref{fig:spectr2},\ref{fig:spectr3}) and
constitutes an antiferromagnetic resonance mode. The relative
difference in resonance fields measured for different orientation
of the sample in this field range is much larger than the relative
change in $g$-factors in the paramagnetic phase. This essential
anisotropy marks the spontaneous symmetry breaking due to the
magnetic ordering above $H_c$.

No theory of the antiferromagnetic resonance in the field-induced
antiferromagnetic phase is developed yet. The theory should
consider the unsaturated order parameter which is induced by the
magnetic field. Possible longitudinal oscillations of the order
parameter should be taken into account.

  We propose here a simplified
treatment of the observed mode based on a molecular-field model,
see, e.g. \cite{kubo}. This model appeared to be adequate for the
description of spin resonance modes in most conventional
antiferromagnets, as well as in impurity-induced ordered phases of
spin-gap magnets \cite{GlazkovPRB,SmirnovPRB}. According to this
model, antiferromagnetic resonance modes correspond to
oscillations of the order parameter and magnetization, which are
affected by the crystal field anisotropy and external field. The
antiferromagnetic resonance frequencies are determined by the
exchange energy and by the anisotropy terms of the second order in
the series expansion of the energy. For a monoclinic crystal the
second-order terms have orthorhombic (two-axes) symmetry
\cite{Vasiliev}. One of the symmetry axes of the second order
anisotropy terms has to be aligned along the two-fold axis $b$. As
at the above analysis of triplet ESR, we chose another axis along
the easy-spin direction caused by the magnetic field in
$b$-direction -- at an angle of 39$^\circ$ from $a$-axis in the
$a-c$ plane.

Thus, at $\mathbf{H}||b$ magnetic field lies along an anisotropy
axis. For $\mathbf{H}\perp(10\overline{2})$  magnetic field is
oriented at an angle of 77.6$^\circ$ with respect to the second
anisotropy axis, and for  $ \mathbf{H}||[201]$ -- at an angle of
12.4$^\circ$. These angles are close to  90$^\circ$ and 0$^\circ$,
respectively, allowing an approximate description of the resonance
spectra by the relations derived for exact orientations. There are
two branches of the antiferromagnetic resonance absorption
\cite{kubo}. In the high-field range above the spin-flop
transition, the first branch has the frequency, approaching to the
paramagnetic resonance frequency, and the frequency of the second
branch does not depend on the field (under the assumption of
constant sublattice magnetization):

\begin{itemize}
  \item[] for $\mathbf{H}$ parallel to the hard axis of anisotropy
\begin{eqnarray}
  f_1&=&\sqrt{\left(\gamma H\right)^2+C_2^2}\\
  f_2&=&C_1\label{eqn:f1}
\end{eqnarray}

  \item[] for $\mathbf{H}$ parallel to the second-hard axis of anisotropy
\begin{eqnarray}
  f_1&=&\sqrt{\left(\gamma H\right)^2+C_1^2}\\
  f_2&=&C_2\label{eqn:f2}
\end{eqnarray}
  \item[] for $\mathbf{H}$ parallel to the easy axis and above the spin-flop field
\begin{eqnarray}
  f_1&=&\sqrt{\left(\gamma H\right)^2-C_1^2}\\
  f_2&=&\sqrt{C_2^2-C_1^2}\label{eqn:f3}
\end{eqnarray}
\end{itemize}

here $C_2>C_1$, $C_i=\sqrt{2H_{Ai}H_E}/\gamma$, $\gamma$ is
magnetomechanical ratio, $H_{Ai}$ and $H_E$ are two effective
anysotropy fields and exchange field respectively.  Each of the
values $H_{Ai}$ and $H_E$ is proportional to sublattice
magnetization, hence to the order parameter.

The branch $f_1$ has the frequency beyond our range 80 GHz in
fields  above 50 kOe. Thus we conclude that the  resonance
frequency in the high-field phase of TlCuCl$_3$ corresponds to the
branch $f_2$ of a two-axes antiferromagnet. The observed
field-dependence of resonance frequencies should be ascribed to
the specific nature of the field-induced ordering, when the order
parameter is induced by the magnetic field and rises from zero at
$H=H_c$. Hence we suppose the resonance frequency observed at
$H>H_c$ is a measure of the field-induced order parameter.

This treatment suggests a universal dependence of  normalized
resonance frequency on the normalized field. The universality
takes place indeed: as shown on the  Fig.\ref{fig:afmr-scale},
three normalized frequency-field dependences taken for different
field orientations coincide well. The normalization factors for
resonance frequency are: 1.0 for $\mathbf{H}||b$, 1.25 for the
$\mathbf{H}||[201]$ and 2.6 for the
$\mathbf{H}\perp(10\overline{2})$.

 The measured antiferromagnetic resonance frequencies follow a square root dependence
 on the magnetic field (solid line on the Fig.\ref{fig:afmr-scale}):
\begin{equation}\label{eqn:afmr-fit}
  f=C\sqrt{\frac{H}{H_c}-1}
\end{equation}
The above consideration of the field-induced order parameter is
consistent with the results of the neutron scattering experiments
performed for $\mathbf{H}\parallel b$
\cite{tanaka-order-neutrons}, where the square root field
dependence was also confirmed near $H_c$.

\section{Conclusions.}

The detailed ESR study of the spin-gap compound TlCuCl$_3$
revealed three kinds of magnetic resonance signals.

New kind of magnetic excitations not reported before is the
antiferromagnetic resonance mode in the field-induced ordered
phase. This branch of spin excitations demonstrates nonlinear and
strongly anisotropic spectrum. From the spectrum of this branch,
in the molecular field approximation, we deduce, that the
field-induced order parameter has a universal dependence for all
directions of the magnetic field: $m \propto \sqrt{H-H_c}$.

Second, we have observed direct ESR transitions between the
singlet ground state and excited triplet state. The negative
field-dependent contribution to the singlet-triplet transition
frequency is temperature dependent. The nature of this temperature
dependence is probably due to the thermal renormalization of the
magnon frequencies. The quantitative theory of this
renormalization is to be constructed.

Third, we found magnetic resonance signals of thermally activated
triplet excitations, and measured the crystal-field splitting of
their spin sublevels. The effective energy of the interaction
between the triplet excitations and the crystal field is obtained.
The formation of the exchange narrowed collective resonance mode
of the interacting triplet excitations was observed.

\section{Acknowledgements.}

Authors  acknowledge  to A.Vasil'ev, A.Kolezhuk, M.Zhitomirsky,
V.Marchenko for discussions. The work is supported by the Russian
Foundation for Basic Research Grant No 03-02-16579, INTAS Grant N
04-5890.

\end{document}